\documentclass{article}

\usepackage[backend=biber]{biblatex}
\usepackage{graphicx}
\usepackage{color}
\usepackage{xfrac}
\usepackage{subfigure}
\usepackage{epsfig}
\usepackage{hyperref}
\usepackage[all]{hypcap}
\usepackage[margin=1in]{geometry}

\begin{document}

\title{The COHERENT Collaboration}
\author{G.C. Rich for the COHERENT Collaboration}

\maketitle

The COHERENT Collaboration seeks to make an unambiguous observation of the coherent elastic neutrino-nucleus scattering (CEvNS) process using a suite of three highly-sensitive detector systems at the Spallation Neutron Source (SNS) of Oak Ridge National Laboratory \cite{coherent-arxiv-2015}.
The detectors proposed to comprise the COHERENT effort are a 14.6-kg CsI[Na] crystal scintillator, 15-kg of germanium P-type point contact detectors, and a dual-phase liquid Xenon (LXe) time-projection chamber.

\begin{figure}[ht]
\centering
\subfigure[Nuclear recoil spectra for CEvNS interactions in COHERENT detectors.] {
	\includegraphics[height=1.5in]{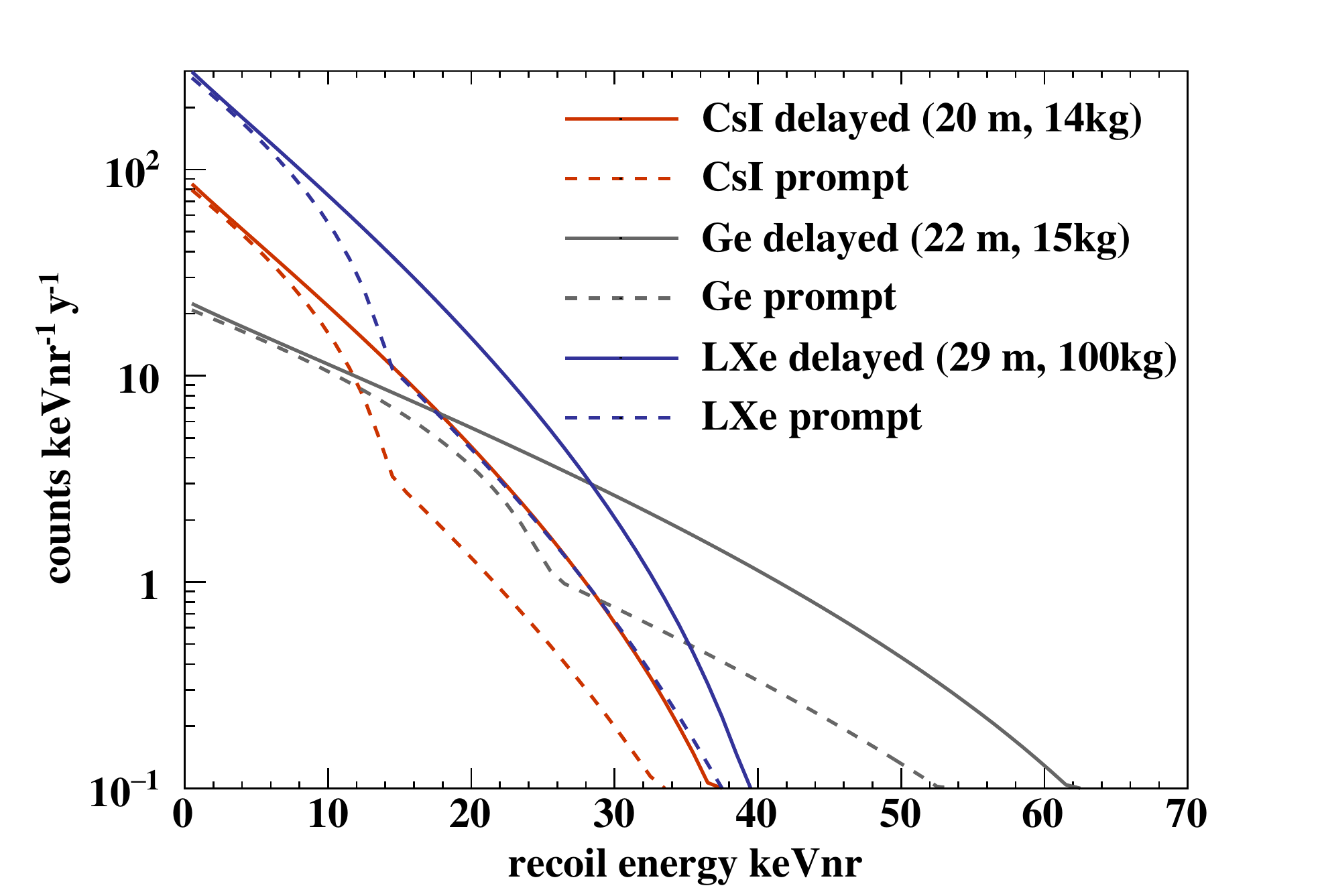}
	\label{fig:COHERENT-a}
	}
\subfigure[Sensitivity for observation of CEvNS by the COHERENT detectors.] {
	\includegraphics[height=1.5in]{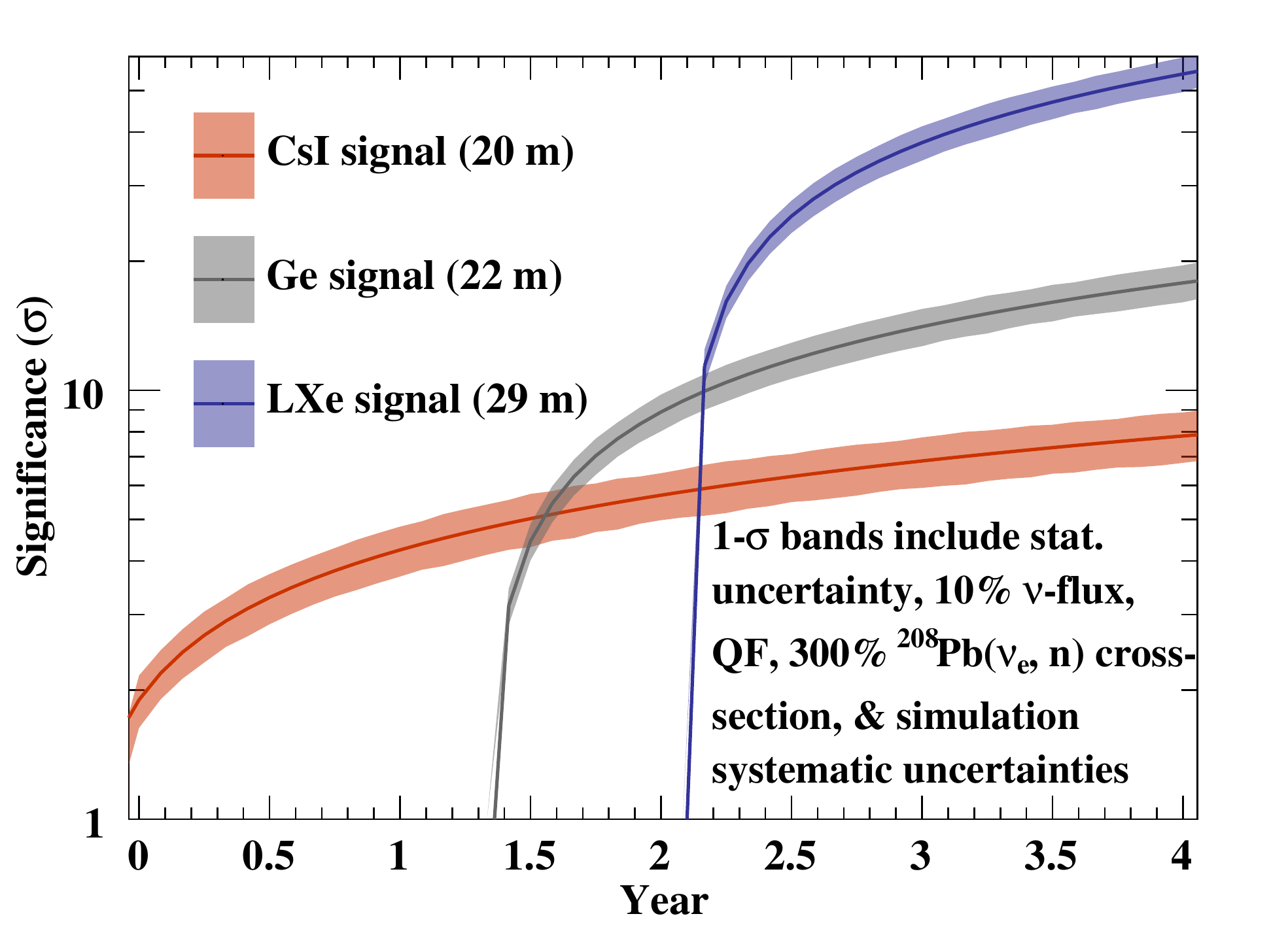}
	\label{fig:COHERENT-b}
	}
\caption{\label{fig:COHERENT} (a) Nuclear recoil spectra for CEvNS interactions at the SNS for the three detector media to be deployed by COHERENT. Recoils associated with the prompt and delayed neutrino populations are shown. (b) Anticipated statistical significance of CEvNS observation by the COHERENT detectors. }
\end{figure}

Stopped-pion neutrino beams, such as the SNS, have been previously been recognized as attractive options for CEvNS searches \cite{Drukier1983,Scholberg2005}. 
The neutrinos produced by the 60-Hz SNS proton beam, fully contained within 700 ns, consist of two populations with distinct energy and timing characteristics: prompt neutrinos, originating from the decay of pions produced in the spallation process, are nearly monoenergetic at 29.9 MeV and are tightly correlated with the beam; the delayed neutrino population is produced by muon decay ($\tau$ = 2.2 $\mu$s) and has neutrino energies up to approximately 50 MeV; figure \ref{fig:COHERENT-a} shows the nuclear recoil spectrum in the COHERENT detectors for SNS neutrinos.
By observing interactions in the delayed-neutrino window, significant reduction of both steady-state and beam-related backgrounds can be realized.
The use of three diverse detector technologies provides an opportunity to reduce sensitivity to detector-related systematic uncertainties and the different nuclear targets in the detectors afford an ability to observe the characteristic $N^2$, neutron-number squared, dependence of the CEvNS cross section. 

In addition to the observation of the CEvNS process, COHERENT will carry out independent, high-precision characterizations of quenching factors (QFs) or detector response for low-energy nuclear recoils and  perform measurements of the cross sections for production of neutrino-induced neutrons (NINs) on various targets, initially including lead and iron.
NINs are a unique background for the COHERENT experiment, with contributions possible from both neutral- and charged-current interactions \cite{elliott2000}, and the process is thought to be involved in nucleosynthesis \cite{qian1997:NINSnucleosynthesis} in addition to being the mechanism by which the HALO supernova neutrino observatory will work \cite{Duba2008}.

Leveraging technological advances and experience from within the low-background physics community, including searches for dark matter and neutrinoless double-beta decay, COHERENT will field detectors capable of observing the low-energy recoiling nuclei produced by CEvNS from the pulsed, tens-of-MeV neutrino beam produced by the SNS, overcoming the significant challenges that have precluded detection of CEvNS for more than 40 years after prediction of its existence \cite{freedman74}.
The anticipated statistical significance of CEvNS observation for COHERENT is shown in Fig. \ref{fig:COHERENT-b}. 
CEvNS data is already being collected by the CsI[Na] detector \cite{collar2014} at the SNS and an extensive background-measurement campaign is underway. 
Relocation of the Xe detector, RED-100 \cite{AAP2015-belov}, to the SNS is expected by mid-2016, and an agreement is in place for loan of the \textsc{MAJORANA} Demonstrator prototype module for the COHERENT effort.
First observation of CEvNS and its $N^2$ behavior will serve to enable future experiments utilizing CEvNS to advance fundamental physics research in addition to informing the direct dark-matter search community about an irreducible background in WIMP detectors \cite{cushman2013:snowmassDM}.

\newpage
\printbibliography

\end{document}